\documentclass[aps,showpacs,preprintnumbers,amsmath,amssymb,nofootinbib,preprint]{revtex4}

\usepackage{graphicx}
\usepackage{color}

\def \be  {\begin{equation}}
\def \ee  {\end{equation}}
\def \ee  {\end{equation}}
\def \bea {\begin{eqnarray}}
\def \eea {\end{eqnarray}}

\def \Tr  {\bf{Tr}}

\begin{document}

\preprint{ECTP-2013-01}

\title{Comment on ''Investigation of Hadron Multiplicity and Hadron Yield Ratios in Heavy-Ion Collisions''}

\author{A.~Tawfik}
\email{a.tawfik@eng.mti.edu.eg}
\email{atawfik@cern.ch}
\affiliation{Egyptian Center for Theoretical Physics (ECTP), MTI University, Cairo, Egypt}

\author{E.~Gamal}
\email{e.gamal@eng.mti.edu.eg}
\affiliation{Egyptian Center for Theoretical Physics (ECTP), MTI University, Cairo, Egypt}

\author{H.~Magdy}
\email{h.magdy@eng.mti.edu.eg}
\affiliation{Egyptian Center for Theoretical Physics (ECTP), MTI University, Cairo, Egypt}

\date{\today}

\begin{abstract}
Oliinychenko, Bugaev and Sorin [arXiv:1204.0103 [hep-ph]] considered the role of conservation laws in discussing possible weaknesses of thermal models which are utilized in describing the hadron multiplicities measured in central nucleus-nucleus collisions. They argued to analyse the criteria for chemical freeze-out and to conclude that none of them were robust. Based on this, they suggested a new chemical freeze-out criterion. They assigned to the entropy per hadron the {\it ad hoc} value  $7.18$ and  supposed to remain unchanged over the whole range of the baryo-chemical potentials. Due to unawareness of recent literature,  the constant entropy per hadron has been discussed in Ref. [Fizika B18 (2009) 141-150, Europhys.Lett. 75 (2006) 420]. Furthermore, it has been shown that the constant entropy per hadron is equivalent to constant entropy normalized to cubic temperature, an earlier criterion for the chemical freeze-out introduced in Ref. [Europhys.Lett. 75 (2006) 420, Nucl.Phys.A764 (2006) 387-392]. In this comment, we list out the ignored literature, compare between the entropy-number density ratio and two criteria of averaged energy per averaged particle number and constant entropy per cubic temperature. All these criteria are confronted to the experimental results. The physics of constant entropy per number density is elaborated. It is concluded that this ratio can't remain constant, especially at large chemical potential related to AGS and SIS energies. 
\end{abstract}. 


\maketitle


\section{Introduction}
In the preprint \cite{sorin}, Oliinychenko, Bugaev and Sorin have considered the role of conservation laws, the values of hard core radii along with the effects of the Lorentz contraction of hadron eigen volumes in discussing the weaknesses of thermal models which are utilized in describing the hadron multiplicities measured in the central nucleus-nucleus collisions. Regardless the unawareness of earlier literature, the authors concluded that none of the criteria for the chemical freeze-out is robust. In doing this, they entirely disregarded the experimental results in baryo-chemical potentials $\mu_b$ and their corresponding temperatures $T$. A systematic analysis of  the four criteria describing the chemical freeze-out is introduced in \cite{Tawfik:2004ss,Tawfik:2005qn,cleymans05}.  Furthermore, a comparison between these four criteria is elaborated in \cite{Tawfik:2004ss,Tawfik:2005qn,cleymans05}. 

Starting from phenomenological observations at SIS energy, it was found that the averaged energy per averaged particle $\langle\epsilon\rangle/\langle n\rangle\approx 1~$GeV \cite{jeanRedlich}, where Boltzmann approximations are applied in calculating  $\langle\epsilon\rangle/\langle n\rangle$, this constant ratio is assumed to describe the whole $T-\mu_b$ diagram. For completeness, we mention that the authors assumed that the pions and rho-mesons get dominant, at high $T$ and small $\mu_b$. The second criterion assumes that total baryon number density $\langle n_b\rangle+\langle n_{\bar{b}}\rangle\approx 0.12~$fm$^{-3}$ \cite{nb01}. In framework of percolation theory, the authors of Ref. \cite{percl} have suggested a third criterion. As shown in Fig. 2 of \cite{Tawfik:2005qn}, the last two criteria seem to give almost identical results. All of them are stemming from phenomenological observation. A fourth criterion based on lattice QCD simulations was introduced in Ref. \cite{Tawfik:2004ss,Tawfik:2005qn}. Accordingly, the entropy normalized to cubic temperature is assumed to remain constant over the whole range of baryo-chemical potentials, which is related to the nucleus-nucleus center-of-mass energies $\sqrt{s_{NN}}$ \cite{cleymans05}. An extensive comparison between constant $\langle\epsilon\rangle/\langle n\rangle$ and constant $s/T^3$ is given in \cite{Tawfik:2004ss,Tawfik:2005qn}. 

The thermodynamic quantities deriving the chemical freeze-out. In framework of hadron resonance gas are deduced \cite{Tawfik:2004ss,Tawfik:2005qn}. Explicit expressions for $s/n$ at vanishing and finite temperature are introduced \cite{Tawfik:2004ss,Tawfik:2005gk}. The motivation of suggesting constant normalized entropy is the comparison to the lattice QCD simulations with two and three flavors. We simply found the $s/T^3=5$ for two flavors and $s/T^3=7$ for three flavors. Furthermore, we confront the hadron resonance gas results to the experimental estimation for the freeze-out parameters, $T$ and $\mu_b$. 

\section{The hadron resonance gas model}
\label{sec:hrg}

The hadron resonances treated as a free gas~\cite{Karsch:2003vd,Karsch:2003zq,Redlich:2004gp,Tawfik:2004sw,Taw3} are
conjectured to add to the thermodynamic pressure in the hadronic phase (below $T_c$). This statement is valid for free as well as strong interactions between the resonances  themselves. 
It has been shown that the thermodynamics of strongly interacting  system can also be approximated to an ideal gas composed of hadron resonances with masses $\le 2~$GeV ~\cite{Tawfik:2004sw,Vunog}. Such a mass cut-off is implemented to avoid the Hagedorn singularity \cite{hgdrn1}. Therefore, the confined phase of QCD, the hadronic phase, is modelled as a non-interacting gas of resonances. The grand canonical partition function reads
\bea
Z(T, V) &=&\Tr\left[ \exp^{-H/T}\right],
\eea
where $H$ is the Hamiltonian of the system and $T$ is the temperature. The Hamiltonian is given by the sum of the kinetic energies of relativistic Fermi and Bose particles. The main motivation of using this Hamiltonian is that it contains all relevant degrees of freedom of confined and  strongly interacting matter. Obviously, it can be characterized by various - but a complete - set of microscopic
states and therefore the physical properties of the quantum systems turn to be accessible in approximation of non-correlated {\it free} hadron resonances. Each of them is  conjectured to add to the overall thermodynamic pressure of the {\it strongly} interacting hadronic matter. It includes implicitly the interactions that result in resonance formation. In addition, it has been shown that this model can submit  a quite satisfactory description of particle production in heavy-ion collisions \cite{Karsch:2003vd,Karsch:2003zq,Redlich:2004gp,Tawfik:2004sw,Taw3}. With the above assumptions the dynamics the partition function can be calculated exactly and be expressed as a sum over 
{\it single-particle partition} functions $Z_i^1$ of all hadrons and their resonances.
\bea \label{eq:lnz1}
\ln Z(T, \mu_i ,V)&=&\sum_i \ln Z^1_i(T,V)=\sum_i\pm \frac{V g_i}{2\pi^2}\int_0^{\infty} k^2 dk \ln\left(1\pm \exp\left[\frac{\mu_i -\varepsilon_i}{T}\right]\right),
\eea
where $\epsilon_i(k)=(k^2+ m_i^2)^{1/2}$ is the $i-$th particle dispersion relation, $g_i$ is
spin-isospin degeneracy factor and $\pm$ stands for bosons and fermions, respectively.

The switching between hadron and quark chemistry is given by the relations between  the {\it hadronic} chemical potentials and the quark constituents; 
$\mu_i =3\, n_b\, \mu_q + n_s\, \mu_S$, where $n_b$($n_s$) being baryon (strange) quantum number. The chemical potential assigned to the light quarks is $\mu_q=(\mu_u+\mu_d)/2$ and the one assigned to strange quark reads $\mu_S=\mu_q-\mu_s$. The strangeness chemical potential $\mu_S$ is
calculated as a function of $T$ and $\mu_i $ under the assumption that the overall
strange quantum number has to remain conserved in heavy-ion collisions~\cite{Tawfik:2004sw}.  

The HRG calculations assume quantum statistics and an overall strangeness conservation. With this regard, the strangeness chemical potential $\mu_S$ is calculated at each value of $T$ and $\mu_b$ assuring that the number of strange particles should be the same as that of the anti-strange particles. It is worthwhile to mention that no statistical fitting has been applied in determining all thermodynamic quantities, including entropy and number density derived from Eq. (\ref{eq:lnz1}).

\section{Physics of constant entropy per number density}
\label{sec:phys}

From the entropy and equilibrium, the Gibbs condition simply leads to 
\bea \label{eq:thrml}
\frac{s}{n} &=& \frac{1}{T} \left(\frac{p}{n}+\frac{\epsilon}{n}-\mu_b\right),
\eea
the rhs is positive as long as  $\mu_b<p/n+\epsilon/n$, where the thermodynamic quantities, $p$, $\epsilon$ and $n$ are supposed to be calculates at the $T-\mu_b$ diagram of the chemical freeze-out. Fig. \ref{fig:tmu1} shows the experimental estimation for the freeze-out parameters 
$T$ and $\mu_b$. It is obvious that increasing $\mu_b$ leads to decreasing $T$ and therefore all values of the thermodynamic quantities decrease as well. Cleymans {\it et al.} \cite{jeanRedlich} suggested an empirical $T-\mu_b$ relation
\bea \label{eq:tmu}
T &=& a-b\, \mu_b^2 - c\, \mu_b^4,
\eea
where $a$, $b$ and $c$ are fitting parameters. In light of this discussion, the value given to $s/n$ can't remain unchanged with increasing  $\mu_b$. Left panel of Fig. \ref{fig:sn1} presents the values of the three criteria $\langle\epsilon\rangle/\langle n\rangle$, $\langle n_b\rangle+\langle n_{\bar{b}}\rangle$ and $s/n$ calculated in HRG, section \ref{sec:hrg}, at $s/T^3=7$. It is obvious that all four criteria seem to remain constant, especially at high $\sqrt{s_{NN}}$. At low energies, the value assigned to $s/n$ \cite{sorin} is larger than the actual one, the value resulted from order conditions. The reason is illustrated in the right panel. At different values for $\mu_b$, the thermal evolution of $s/n$ is presented. It is obvious that  $s/n$ never reaches $7.18$ at $\mu_b>500~$MeV. It is essential to bear in mind that the value $7.18$ has almost no physical interpretation. It is just an {\it ad hoc} value. This makes it inapplicable at Alternating Gradient Synchrotron (AGS) and Schwer-Ionen-Synchrotron (SIS) energies. Almost same kind of restriction would be valid for $\epsilon/n$. According to Eq. (\ref{eq:thrml}), 
\bea
\frac{\epsilon}{n} &=& T \frac{s}{n} + \mu_b -\frac{p}{n}.
\eea

The physics of constant $s/T^3$ has been discussed in Ref. \cite{Tawfik:2004ss,Tawfik:2005qn}. It combines the three thermodynamic quantities, $p/T^4$, $\epsilon/T^4$ and $n/T^3$
\bea
\frac{s}{T^3} &=& \frac{p}{T^4} + \frac{\epsilon}{T^4} - \mu_b\, \frac{n}{T^3}.
\eea
At chemical equilibrium, the particle production at freeze-out is conjuncted to fully fulfil the laws of thermodynamics, as Eq. (\ref{eq:thrml}). The hadronic abundances observed in the final state of heavy-ion collisions are settled when $s/T^3$ drops to $7$ i.e., the degrees of freedom drop to $7 \pi^2/4$. Meanwhile the
changing in the particle number with the changing in the collision energy is given by $\mu_b$, the energy that produces no additional work, i.e. the stage of vanishing free energy, gives the entropy at the chemical equilibrium. At the chemical freeze-out, the equilibrium entropy represents the amount of energy that can’t be used to produce additional work. In this context, the entropy is defined as the degree of sharing and spreading the energy inside the system that is in chemical equilibrium \cite{Tawfik:2005qn}.

\section{Constant Entropy per Number in Lattice QCD Simulations and Heavy-ion Collisions}

Once again, related literature on lattice QCD simulations is not cited in \cite{sorin}. For example, Borsanyi {\it et al.} \cite{fodor12} studied the trajectories of constant $s/n$, where $s=S/V$ and $n=N/V$, on the phase diagram and thermodynamic observables along these isentropic lines. This was not the only work devoted to such line of constant physics \cite{old84}. In Stefan-Boltzmann limit, the ratio $s/n$ is assumed to remain unchanged with increasing $\mu_b$ (Appendix A of \cite{fodor12}). In doing this, lowest order in perturbation theory is assumed, where strangeness chemical potential $\mu_S$ likely vanishes. For $\mu_b/T$, a limiting behavior for the isentropic lines on the phase diagram is obtained. The ratio $s/n$ has been measured at various $\sqrt{s_{NN}}$ \cite{expp}. It is concluded that in limits of low temperatures, increasing the chemical potential results in an overestimation for the ratio $s/n$ even beyond the applicability region of the Taylor-expansion method, which is applied in lattice QCD simulations at finite chemical potential. Two remarks are now in order. First, the values of $s/n$ seem to depend on the chemical potential $\mu_b$ or $\sqrt{s_{NN}}$. This is confirmed in different experiments \cite{expp} and lattice gauge theory \cite{fodor12}. Second, the ratio $s/n$ as calculated in the lattice QCD simulations \cite{fodor12} is suggested to characterized the QCD phase diagram \cite{Tawfik:2004sw}. The QCD phase diagram is likely differs from the freeze-out diagram \cite{Tawfik:2004ss,Tawfik:2005qn}, especially at large chemical potential $\mu_b$ or small $\sqrt{s_{NN}}$ so that at fixed $\mu_b$ the critical temperature differs from the freeze-out temperature.

\section{Results and Conclusions}
\label{sec:resl}

\begin{figure}[htb]
\includegraphics[width=7.cm,angle=-90]{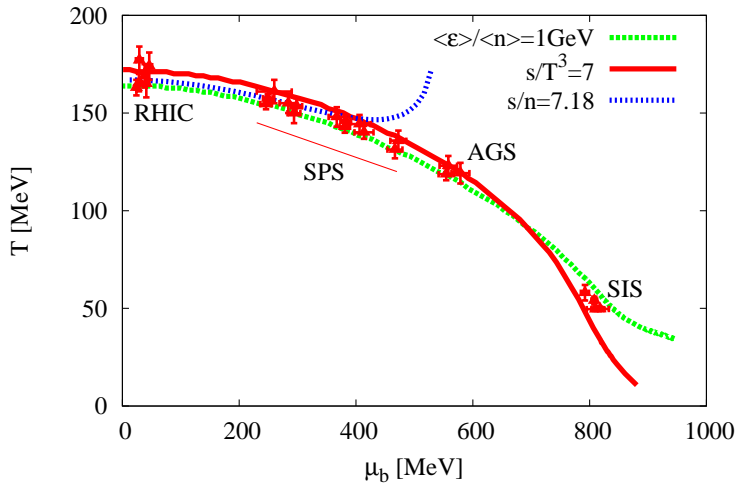}
\caption{The freeze-out parameters, $T$ and $\mu_b$, measured in various heavy-ion collisions experiments (labelled) are compared with the three criteria, $\langle\epsilon\rangle/\langle n\rangle=1~$GeV (dashed line), $s/n=7.18$ (dotted line) and $s/T^3=7$ (solid line). }
\label{fig:tmu1}
\end{figure}

\begin{figure}[htb]
\includegraphics[width=6.cm,angle=-90]{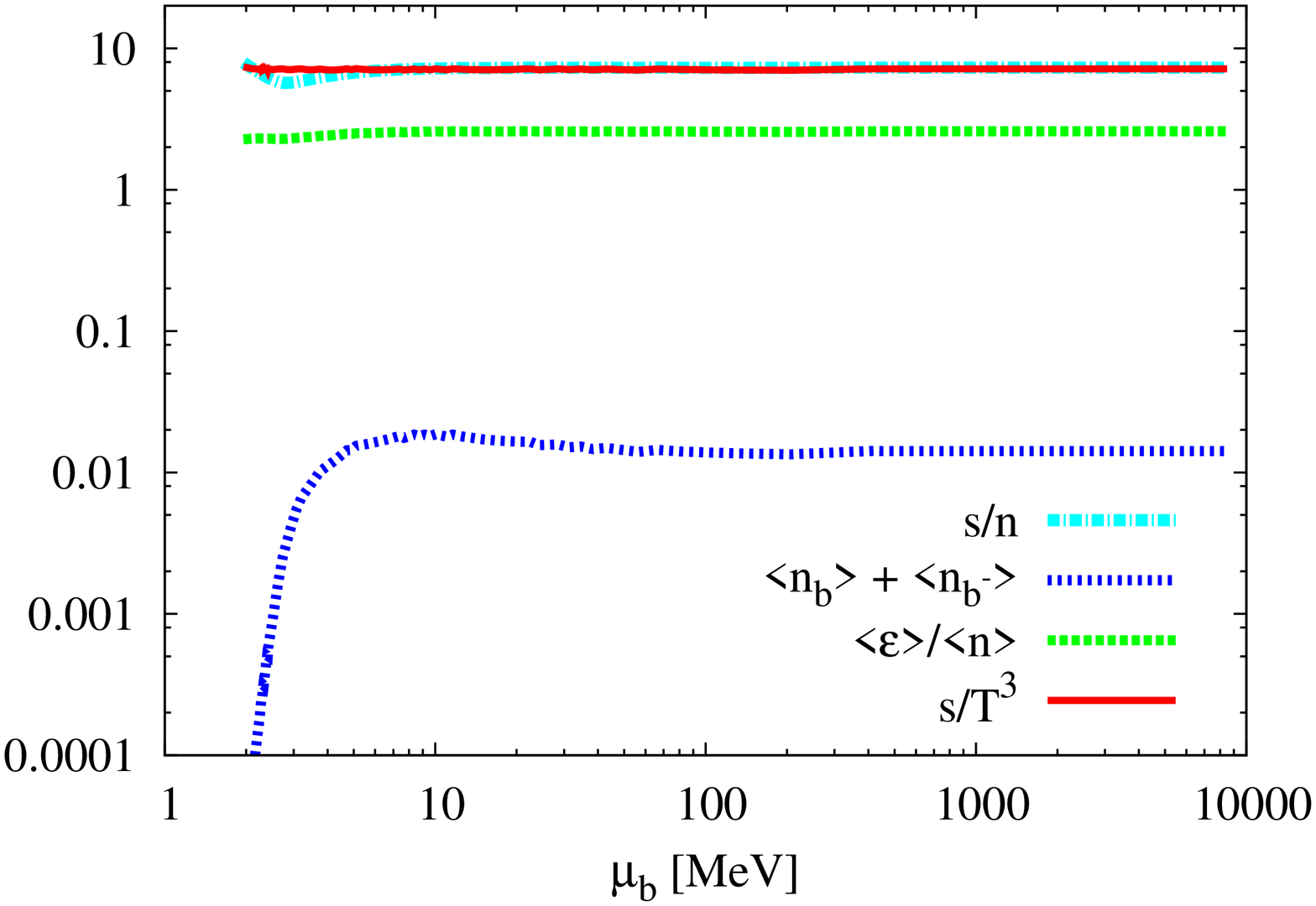}
\includegraphics[width=5.5cm,angle=-90]{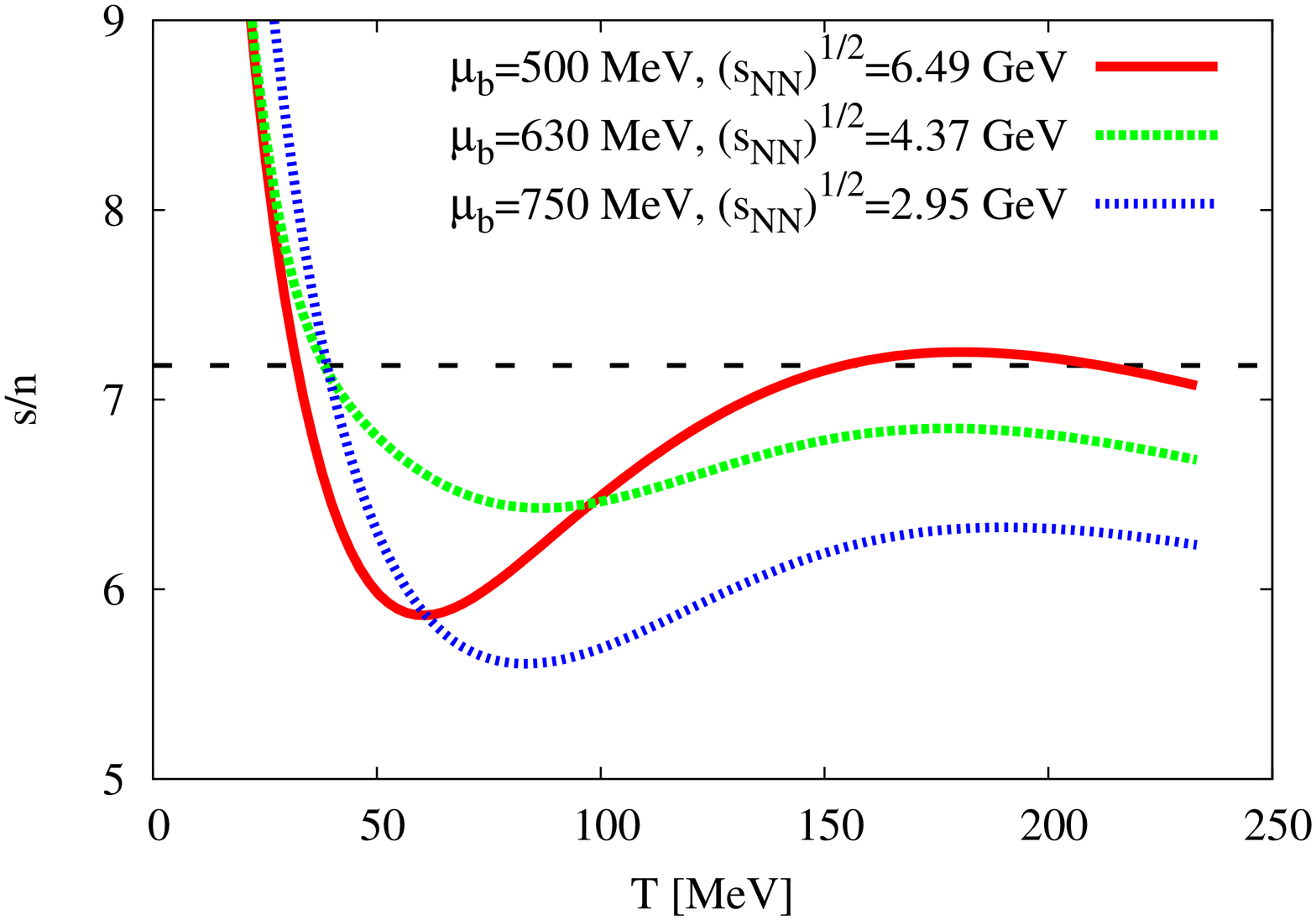}
\caption{Left panel: the energy scan for the three criteria, $\langle\epsilon\rangle/\langle n\rangle$, $\langle n_b\rangle+\langle n_{\bar{b}}\rangle$ and $s/n$ are calculated in HRG, section \ref{sec:hrg} at $s/T^3=7$. Right panel: the thermal evolution of $s/n$ at $\mu_b=500$, $630$ and $750~$MeV is presented. The horizontal dashed line indicates $s/n=7.18$. The singularities at low temperatures are stemming from almost vanishing number density (compare with left panel).}
\label{fig:sn1}
\end{figure}

In Fig. \ref{fig:tmu1}, the freeze-out parameters, $T$ and $\mu_b$, measured in various heavy-ion collisions experiments are compared with the three criteria, $\langle\epsilon\rangle/\langle n\rangle=1~$GeV (dashed line), $s/n=7.18$ (dotted line) and $s/T^3=7$ (solid line). The experimental data are taken from \cite{cleymans05} and the reference therein. The quality of each criterion is apparent. All conditions are almost equivalent at very high energy. The ability of the condition $\langle\epsilon\rangle/\langle n\rangle=1~$GeV at very low energies are not as much as that of $s/T^3=7$. As discussed in section \ref{sec:phys}, $s/n=7.18$ seems to fail to reproduce the freeze-out parameters at $\mu_b>500~$MeV. To illustrate the reason for this observation, the thermal evolution of $s/n$ at very high chemical potential calculated in HRG is presented in the right panel of Fig. \ref{fig:sn1}. Details on HRG are elaborated in section \ref{sec:hrg}. It is obvious that the value assigned to $s/n$ would be achieved at $\mu_b>500~$MeV. In other words, it is obvios that the behavior of $s/n$ is non-monotonic.

The left panel of Fig. \ref{fig:sn1} presents the energy scan for the three criteria, $\langle\epsilon\rangle/\langle n\rangle$, $\langle n_b\rangle+\langle n_{\bar{b}}\rangle$ and $s/n$  calculated in HRG, section \ref{sec:hrg} at $s/T^3=7$. The calculations in HRG are performed as follows. Starting with a certain $\mu_b$, the temperature is increased very slowly. At this value of $\mu_b$ and at each increase in $T$, the strangeness chemical potential $\mu_S$ is determined to assure strangeness conservation. Having the three values of $\mu_b$, $T$  and $\mu_S$, then all thermodynamic quantities are calculated. When the ratio $s/T^3$ reaches the value $7$, then the three quantities $\langle\epsilon\rangle/\langle n\rangle$, $\langle n_b\rangle+\langle n_{\bar{b}}\rangle$ and $s/n$ are registered. This procedure is repeated over all values of $\mu_b$. We find that $s/T^3=7$ assures $s/n=7.18$ and $\langle\epsilon\rangle/\langle n\rangle=1~$GeV at small $\mu_b$ (large $\sqrt{s_{NN}}$). At large $\mu_b$ (small $\sqrt{s_{NN}}$), the values of $s/n$ gets smaller values, so that the applicability of $s/n=7.18$ is limited to $\mu_b<500~$MeV. In conclusion, the robustness of  $s/n=7.18$ is very much limited in comparison to the four criteria: percolation \cite{percl}, baryon number \cite{nb01}, energy per particle \cite{jeanRedlich} and normalized entropy \cite{Tawfik:2004ss,Tawfik:2005qn}. That $s/T^3$ is accompanied with constant $s/n$ has been introduced in Ref. \cite{Tawfik:2004ss}. That authors of \cite{sorin} argue that $s/n=7.18$ is novel likely reflects an ignorance of related literature. The four criteria \cite{Tawfik:2004ss,Tawfik:2005qn,jeanRedlich,nb01,percl} are based on physical observation either phenomenological and/or theoretical. The authors of \cite{sorin} are suggesting an {\it ad hoc} value for the ratio $s/n$. It is inapplicable at AGS and SIS energies. Its relation to $s/T^3$ is apparently overseen. The same is valid for the comparison with other criteria (some of them are ignored, completely) and ignorance of the experimental measurements. The {\it ad hoc} value assigned to $s/n$ is obviously much robuster than any other criterion. Unawareness of literature and underestimating or even ignoring previous work are violation of rules of the scientific research.

\end{document}